\documentclass[12pt]{article}

\usepackage{amsmath}
\usepackage{epsfig}

\def\r{{\mathbf{r}}}

\def\be{\begin{equation}}
\def\ee{\end{equation}}
\def\eq#1{{Eq.(\ref{#1})}}

\def\r{{\bf r}}
\def\be{\begin{equation}}  \def\ee{\end{equation}}

\begin{document}

\renewcommand{\thefootnote}{\fnsymbol{footnote}}

\begin{titlepage}

\begin{center}
\vspace*{.8 in} {\Large\bf On the formation of caveolae and similar membrane invaginations}\\ \vspace*{0.6 in} {\large Pierre Sens$^{1}{}^\dagger$ and Matthew S. Turner$^{2}$}\\ \vspace*{0.6 in} {\large December 12, 2002}
\\ \vspace*{0.6 in} {$^{1}$Institut Charles Sadron, 6 rue Boussingault, 67083 Strasbourg
Cedex, France} 
\\ \vspace*{0.2 in} {$^{2}$Department of Physics, University of Warwick, Coventry CV4
7AL, UK}
\\ \vspace*{0.2 in} {\small $\dagger$ sens@ics.u-strasbg.fr}
\\ \vspace*{0.2 in} {\small Classification: Physical Sciences - Physics}
\end{center}

\end{titlepage}

\noindent {\bf ABSTRACT \hspace{0.1in} We study a physical model for the formation of bud-like invaginations on
fluid membranes under tension, and apply this model to caveolae formation. We demonstrate that budding can be driven
by membrane-bound inclusions (proteins) provided that they exert
asymmetric forces on the membrane that give rise to bending
moments. In particular, Caveolae formation may not necessarily require
forces to be applied by the cytoskeleton. Our theoretical model is able to explain several
features observed experimentally in caveolae, where proteins in the
caveolin family are known to play a crucial role in the formation of
caveolae buds. These include (i) the formation of caveolae buds with sizes in the 100nm range (ii)
that a fairly large variation of bud shape is expected (iii)
that certain N and C termini deletion mutants result in vesicles that are
an order of magnitude larger. Finally, we discuss the possible origin of
the morphological striations that are observed on the surfaces of the caveolae.}

\vspace{0.5cm}
\section{Introduction}\label{introduction}

It has long been understood that invaginations form spontaneously on cell
membranes\cite{alberts}. These
invaginations, which eventually separate from the membrane as mature, membrane bound vesicles, play an essential role in cellular trafficking and
signalling\cite{review3,lisantiJCB94}. The mechanism by which such
invagination is controlled is still far from fully understood,  although it
is now widely accepted that certain membrane-bound proteins, including clathrin and
caveolin, play an important role. The formation of clathrin coated pits is thought to be
driven by the controlled geometric aggregation of clathrin into rigid scaffolding which force the membrane to curve\cite{clathrin1,mashl}. The mechanism for formation of the second most common class of membrane invaginations, known as caveolae, is less well understood. Caveolae, which are less morphologically
distinct than clathrin coated pits, resemble $\Omega$-shaped invaginations with a typical size of the order
of 100nm\cite{rothberg,review1,structure1}. They
are present at high concentrations on primary adipocytes, fibroblasts, muscle cells, pulmonary type 1 cells as well as endothelial cells, and perform a variety of function ranging from signal transduction to intracellular transport\cite{function1,cterm1}. A ``striated coat'' can be seen on the cytoplasmic side of the caveolae membrane. It is believed to reflect the organization of a recently discovered class of membrane-bound
proteins, called caveolins, which are crucial to the formation of caveolae\cite{lisantiJCB94}.

The protein caveolin has an hairpin structure, with a short membrane spanning
sequence, flanked by two hydrophilic termini, both found on the cytoplasmic
side of the cell membrane: a 101 amino-acid polypeptide N-terminus tail (polymer), and a shorter (44 a-a) C-terminal, which is strongly attached to the membrane\cite{cterm1}. These caveolin molecules are typically found in small
aggregates of
15-17 molecules\cite{review1,oligo}, the aggregation being driven by residues of the N-terminal located close to the membrane. Furthermore it is believed\cite{cterm1} that there exist some specific C-terminal - C-terminal attractions, which are responsible for the organization of the protein aggregates at the surface of the caveolae membrane. Mutational analysis of caveolin-induced vesicle formation have been recently performed\cite{mutation} and are discussed in relation with our theory in
section \ref{concs} below.

Caveolae are now thought to influence cell physiology in many ways,
including growth and cell division, adhesion, and hormonal
response\cite{choles1}. These invaginations  have been associated with the formation of lipid rafts\cite{raft1},
glycosphingolipid- and cholesterol-enriched microdomains within the plasma
membrane of eucaryotic cells.  Their ability to perform many different tasks might be achieved by their involvement in reporting change in membrane composition by signal transduction to the nucleus. It may also be connected to their regulation of signal traffic in response to extracellular stimuli, including mechanical stress\cite{shear}.  

From a physical point of view, spontaneous vesicle formation has been observed in vitro by adding amphiphilic polymers to various lipid systems\cite{vesicle}. It can be viewed as an example of the so-called curvature instability of fluid membranes containing inclusions, predicted to occur for inclusions that locally influence the membrane curvature\cite{leibler}. There have been physical studies of the inclusion-induced budding of vesicles\cite{kim,seifert,lipowsky3} and works on the effect of single\cite{lipowsky1} and distributed\cite{fourcade} polymers grafted on
membranes. 

Our aim is to study the effect of small inclusions,
such as proteins, that affect the shape of the cell membrane. We assume that this ``foreign'' object exerts a force on the membrane, which may be due either to entropic effects, similar in origin to the
pressure exerted by a gas onto the walls of its container, or to
specific mechano-chemical forces. Throughout we will attempt
to compare our rather general theory with the specific phenomenon of
caveolin-mediated formation of caveolae. The fact that membrane-bound objects exert a force on the membrane arises naturally from theories that describe polymers grafted to surfaces. These have
been extensively developed over the last decade or so, based on early ideas due to de Gennes\cite{degennes} and others. The forces exerted by membrane-bound inclusions, as well as there interactions, have been calculated in certain ideal situations (e.g. idealized polymers on a tensionless membrane)\cite{carlos,lipowsky2}. In what follows, we will analyzed arbitrary force distributions, which allow for the description of specific inclusions, such as the caveolin aggregates. We also focus on tension bearing membranes, a situation that we believe more closely approximates the plasma membrane of the cell.

The deformation of a membrane subjected to an arbitrary force distribution, and the subsequent membrane-mediated interaction between two such  distributions, are computed in sec.\ref{analysis}. The description of possible physical models for the force distribution follows in sec.\ref{models}. A physical theory for the several levels of protein self-organization at the cell membrane is presented in section \ref{buds}, followed by our results for different force distributions (sec.\ref{results}). We then briefly comment on possible physical mechanisms for the peculiar phase behavior (stripe formation) at the bud surface (sec.\ref{stripes}), and conclude in section \ref{concs}.

\section{Membrane response to an arbitrary force distribution}\label{analysis}

The deformation energy of a membrane involves its surface tension $\gamma$ and bending rigidity $\kappa$. Cells commonly adjust their surface tension to a set
value via a mechanism known as  Surface-Area-Regulation\cite{SAR}. Hence membrane
phenomena over  sufficiently long timescales effectively occur at constant
surface  tension. It is also known that the composition of biological membranes exhibits spatial
variations. Caveolar membranes for instance, show a high cholesterol content\cite{choles1}, the precise biochemical role of which is not yet entirely clear. From a physical point of view, it is known that cholesterol increases the local rigidity of the membrane\cite{kapparef,cholesterolkappa}. Local variations of membrane rigidity are not included in the following  model, but some (limited) information on the impact of cholesterol on caveolae at the physical level can be obtained by examining the effect on {\em uniform} changes in $\kappa$ across the whole membrane. 

Initially, we restrict our analysis to a membrane that is weakly deformed by the presence of the inclusions. We proceed by writing down the free energy of an infinite
planar fluid membrane as a standard expansion in powers and gradients of the membrane displacement $u(\r)$ from its flat (unperturbed) position\cite{safran}. This is modified to include the leading order term arising from the coupling to $u$ of the applied pressure distribution $f(\r)$, which will arise from the action
of inclusion(s)\cite{nonetforce}:
\be\label{freeenergy1}
     {\mathcal F}=\int d^{2} \r' \biggl[
     \frac{\kappa}{2}(\nabla^{2} u)^{2} +
\frac{\gamma}
     {2}(\nabla u)^{2}  - fu\biggr]
\ee
Typically values for phospholipid bilayers are\cite{kapparef} $\kappa\approx 20
k_{B}T$\footnote{It is usually helpful to compare energies to the energy available from thermal fluctuations $k_BT$, where $k_B$ is the Boltzmann constant and $T$ the temperature}. The surface tension is reported to be in the range\cite{sheetz}
$\gamma\approx 10^{-2}-10^{-1}$ pN/nm. The interplay of surface tension and bending rigidity defines a characteristic lengthscale $k^{-1}\equiv\sqrt{\kappa/\gamma}\sim 30-90nm$. 

Minimization of this energy results in the equilibrium membrane displacement $u(\r)$, and is reported in more detail elsewhere\cite{multipolepaper}. We find $u(\r)=\int G(\r-\r') f(\r')
d^2\r'$ with the Green's function (the response to a point force) given by
\be
     G(\r-\r') = -\frac{1}{2\pi\gamma}
     \biggl[K_{0}(k|\r - \r'|) + \log{k|\r - \r'|}
     \biggr]\label{realspacegreens}
\ee
where $K_0$ is a modified Bessel function\cite{bessel}, which decreases exponentially over a size $k^{-1}$. The membrane displacement is discussed further in section \ref{models}.  

Overlap of displacements due to neighboring inclusions lead to membrane-mediated interactions between them. The interaction potential $\Phi({\r})$ between two similar inclusions separated by a vector ${\bf r}$ is obtained by inserting the total force distribution $f(\r')+f(\r'+\r)$ into \eq{freeenergy1} and identifying the $\r$-dependence of the resulting energy (see \cite{multipolepaper} for the general theory). The interaction energy reduces to
\be
     \Phi(\r) = -\int d^2\r' \int d^2\r''
     f(\r') f(\r'') G(\r -\r' +\r'')
\ee
where $G(\r)$ is the real space Green's function given by
\eq{realspacegreens}.

If the inclusions have a circular symmetry ($f(\r')=f(r')$) and do not overlap ($r>2b$ where $b$ is the spatial extent of the force), we have
been able to determine the interaction potential
exactly in an analytic form
\be
\Phi(\r) = \frac{1}{2\pi\gamma} \zeta^2 K_{0}(kr)\label{phisymm}
\ee
where $\zeta= \int_0^\infty 2\pi r'dr'\psi(r')I_0(kr')$
characterizes the strength of the interaction ($I_0$ is another modified Bessel function\cite{bessel}). The
interaction is everywhere repulsive in the regime of validity $r>2b$.

\section{Models for membrane-bound proteins}\label{models}
Up to this point, we have been able to avoid making any but a few rather general assumptions about the form of the force exerted by the membrane inclusions. We will now proceed to consider some specific models for the force distribution. We do this both to make possible the later quantitative comparison with experiments and to demonstrate how such forces are expected to arise on general physical grounds. The force distributions and the subsequent membrane deformations (from \eq{realspacegreens}) are shown in Fig.1.

\subsection{Flexible polymers}

In this section we treat caveolin proteins as flexible, linear polymer
chains, anchored to the membrane. The idealized picture enable us to extract
an analytic estimate of the force distribution. In the language
of polymer physics\cite{degennes}, the caveolin homo-oligomer can be viewed as a brush of $Q\simeq 16$ polymer chains, grafted by one end to a small patch of
membrane of radius $a$. The flexible chains on average arrange
themselves radially to form a hemisphere of radius
$b$ (Fig.1a).
Thus for radial distances $a<r<b$ one finds a corona of
randomly coiled polymer chains with a chain density that is larger near
the core and smallest on the outskirts of the distribution. 

A central concept in the theory of polymer
physics is the existence of a correlation length or `blob size'
$\xi(r)$\cite{degennes}, which is roughly the distance between interchain contacts in the
corona of the caveolin brush (viewed as a semi-dilute polymer solution). Each chain can then be thought of
as a string of correlation blobs extending radially outwards, with small values of
$\xi$ corresponding to large densities of monomers. The classical Daoud-Cotton model\cite{daoudcotton} takes advantage of the fact that the surface area of a
hemisphere of radius $r$ is approximately filled by $Q$ close-packed
blobs, to deduce the scaling of the
correlation length: $\xi(r) =r \sqrt{2\pi/Q}$. It can be shown\cite{degennes} that the work done in generating
each blob is $k_{B}T$, independent of the blob size. Thus we may write the
pressure in this region as the energy per blob
divided by the volume of a blob
\be
f(\r)=-\frac{k_{B}T}{\xi(r)^3} =-
\left(\frac{Q}{2\pi}\right)^\frac{3}{2}
\frac{k_{B}T}{r^3}
\label{powerlaw}
\ee
A result that is valid for $a<r<b$, and that is consistent with more detailed calculations\cite{carlos,lipowsky2}. The physical origin of this pressure
can be understood as simply due to the exchange of momentum due to
collisions of the polymer chains with the membrane. It is therefore purely
entropic in origin, as evidenced by the overall $k_{B}T$ energy scale. For
$r>b$ the pressure is zero since the chains have finite length (in fact, it is exponentially small\cite{carlos,lipowsky2}). The
pressure in the core binding region is assumed constant and must involve a
total force equal and opposite to that applied by the corona\cite{nonetforce}
\be
f(r) =\frac{E_0}{a^3}
\left\{\begin{array}{ll}       
1& {\rm if} \quad 0<r<a \\
-\frac{a^3/r^3}{2(1-a/b)} & {\rm if} \quad a<r<b \\
 0 & {\rm if} \quad r>b
\end{array}\right. \; 
\label{cc}
\ee
Note that the strength of the force is characterize by the energy scale $E_0=f(r=0)a^3$. For the Q-chain oligomer, it is $E_0=2 k_BT (Q/2\pi)^{3/2}(1-a/b)\sim4k_BT$.

\subsection{Block Distribution}

We believe that the polymer brush model captures some of the fundamental properties of a collection of large hydrophilic proteins anchored to a biomembrane, namely {\em i)} a downward pressure exerted by the cytosolic portion of the proteins, combined with {\em ii)} an upward pull from the anchors (the hydrophobic region of the proteins). It however employs rather strong assumptions (random coil configuration, absence of internal structure, and large size of the polymer chains) which are certainly not satisfied for the protein caveolin. The simplest example of a general force distribution that satisfies the criteria above is a ``block distribution'', for which
the force exerted by the hydrophobic anchors (between $0<r<a$) and the hydrophilic sections ($a<r<b$) are both constant:  $E_0/a^3=f_{r<a}=(b^2/a^2-1)f_{a<r<b}$. The membrane deformation for such distribution is larger than for the brush distribution for the same strength, as characterized by the energy $E_0$, since the force is not concentrated near the center of the distribution (Fig.1c). 
\section{bud structure and morphology}\label{buds}

 \def\sp{{p^*}}		\def\Osp{\Omega_\sp}

Caveolae formation involves a hierarchy of self organization, ranging from the nanometric scale (oligomers of $\sim15$ particles and size $\sim5\ nm$) up to  buds of radius $\sim100\ nm$. We review briefly the theoretical framework of thermal self-organization\cite{safran}, and give insights on the caveolin homo-oligomer formation, which we view essentially as a micellization in two dimensions. We then describe in some details the formation of caveolae buds. 

Consider a solution of particles of average surface fraction $\bar\phi$, that can exist either as isolated entities or in larger aggregates (homo-oligomers) of $p$ particles and of energy$F_p=p f_p$. The concentration $C_p$ of p-sized aggregates follows a Boltzmann law\cite{safran}: $C_p\sim e^{-(F_p-\mu p)/k_BT}$, where $\mu$ is the chemical potential of the particles, usually fixed by the average concentration $\bar\phi$. There is an energetic tendency to form aggregates if the energy per particle $f_p$ is (at least in some regime) a decreasing function of the aggregation
number $p$.  It overcomes the entropic dispersive effect beyond a critical value of $\bar\phi$ (the critical aggregation concentration, or c-a-c), usually defined as the concentration at which the density of aggregates is equal to the density of isolated particles. At the c-a-c and above the average size $\sp$ of the aggregates is the one that minimizes the energy per protein in the aggregate. The root mean squared deviation $\Delta p$ from the average depends upon the steepness of the energy variation around that minimum. Expressed in the form of equation, these conditions yield (for $\sp\gg1$):
\be
f_\sp-\mu_{cac}={\partial f_p\over \partial p}|_\sp=0\qquad \Delta p=\sqrt{k_BT\over\partial^2_pF_p}|_\sp
\label{selfag}
\ee

The driving force for homo-oligomerization is an attractive interaction between specific motives on the N-terminal of the protein\cite{oligo}. We proceed by assuming that all proteins in the interior of the oligomer experience a mutual attraction, and contribute  to the oligomer energy $F_p$ via a negative linear term $-\mu' p$. Proteins in the outskirts of the oligomer on average experience less attraction, as they have less neighbors. They increase the energy $F_p$ by a factor $+\pi \beta \sqrt{p}$, where $\beta$ is the energy loss ({\em per protein}) for being on the outskirts\footnote{the equivalent of the surface tension of a liquid}. Bringing proteins together also leads to steric and/or entropic repulsion (crowding). For a polymer brush for instance, the latter contribution to $F_p$ is $+\alpha p^{3/2}$ (see \eq{powerlaw}). Solving \eq{selfag} with $F_p/p\equiv f_p=\alpha\sqrt{p}+\pi \beta/\sqrt{p}-\mu'$, the experimental observation (oligomers containing $14-16$ proteins\cite{oligo}) are consistent with $\alpha=2k_BT$ and $\beta=10k_BT$. Both numbers are physically reasonable: the frustration $\alpha$ which results in bringing many proteins close to one another might be expected to be of order $k_BT$ per protein, and the calculated energy loss $\beta$ for proteins on the outskirts of the oligomers is of order an hydrogen bond energy ($10 k_BT$) per proteins.  Although quite crude, this model provides a thermodynamic description for  the first level of self organization in caveolar membranes, which is able to reproduce the experimental observation on caveolin oligomerization, namely $\sp\simeq 15$ and $\Delta p\simeq 2$. 

The formation of the caveolae themselves can be described more accurately. As it involves a large number of oligomers, a precise description at the molecular level is less crucial. We model the $\Omega$-shaped invagination by a closed sphere of radius $R$. The bending moments exerted by the protein oligomers, which are all on the cytoplasmic side of the membrane, drives the bud formation, expected to occur above a critical budding concentration (c-b-c) of oligomers (see Fig.2). 
For small concentration $\phi<\phi_{cbc}$ (Fig.2a), the membrane is uniformly covered by oligomers and remains almost flat. Buds start forming as the concentration increases, and outnumber isolated oligomers at the c-b-c (Fig.2b). If the concentration is increased further (Fig.2c), then the concentration of isolated oligomers, and the bud size, remain almost constant ($\phi_1=\phi_{cbc}$), while the number of buds increases.

The free energy per membrane inclusion in the bud $f_b$ (\eq{enbud} below) contains several contributions. Energy is gained if the membrane curves to accommodate the deformation imprinted by the caveolin oligomer. A membrane curving away from the caveolin aggregate is favored (first term, RHS of Eq.(\ref{enbud})). In the limit of small
curvature, the energy reduction per oligomer is of order $-D_{rr}/R$ (Eq.(\ref{pio6})). On the
other hand,  bud formation costs an energy that depends on the bending rigidity and
surface tension (second and third terms, RHS of \eq{enbud}), and leads to higher local oligomer
concentration, modifying the pair interaction energy (fourth term, RHS of \eq{enbud}). This interaction is characterized
by the second virial coefficient $B_2$ (\eq{pio6}), and involves the full interaction potential $V(r)$ (see sec.\ref{results}). The last term in the RHS of \eq{enbud} is the mixing entropy of a gas of membrane inclusions on a lattice.
\begin{eqnarray}
f_b=-{D_{rr}\over R}+2{\kappa s_1\over\phi
R^2}+{\gamma s_1\over\phi}+\phi{k_BT B_2\over
s_1}\cr+k_BT\left(\log\phi+({1\over\phi}-1)\log(1-\phi)\right)
\label{enbud}
\end{eqnarray}
where $s_1=\pi b^2$ is the oligomer area, and with
\be
D_{rr}\equiv \int \frac{d^2r}{2}r^2 f(r)\quad
B_2\equiv\int
\frac{d^2r}{2}\left(1-e^\frac{-V(r)}{k_BT}\right)
\label{pio6}
\ee

Minimizing this energy with respect to $R$ gives the optimal bud radius $R^*=R_{min}/\phi$, where $R_{min}\equiv 4\kappa s_1/D_{rr}$  corresponds to the minimal radius of a bud densely packed with caveolin. Further energy minimization leads to the optimal amount of protein $\phi^*$ recruited in the bud, defined by:
\be
\frac{\log{(1-\phi^*)}}{\phi^*{}^2}+\frac{\gamma s_1}{k_BT \phi^*{}^2}={B_2\over s_1}-{2\kappa s_1\over
k_BTR_{min}^2}
\label{pio7}
\ee
This equation has a clear physical meaning. The protein coupling to the membrane curvature effectively reduces the second virial coefficient by an amount $B_2^{eff}\equiv2\kappa s_1^2/(k_BTR_{min}^2)$, which indicates an attraction between oligomers\cite{leibler}. 

The optimal concentration of \eq{pio7} corresponds to an energy minimum $f_b^*$. At the c-b-c (\eq{selfag}), it is equal to the oligomer chemical potential: $f_b^*=\mu_{cbc}$. The latter can be related to the concentration $\phi_1$ of isolated oligomer via $\mu=\log(\phi_1/(1-\phi_1))+2 B_2\phi_1$, which is the chemical potential of a gas on a lattice with pair interaction. The equation defining the critical budding concentration is:

\begin{eqnarray}
\log{\phi_{cbc}\over 1-\phi_{cbc}}+2 {B_2\over s_1}\phi_{cbc}=\cr
\log{\phi^*\over 1-\phi^*}+2\frac{B_2-B_2^{eff}}{s_1}\phi^*
\label{cbc}
\end{eqnarray}

The mean variation of the radius $\Delta R^2\equiv\langle R^2\rangle-R^*{}^2$ can be approximately calculated by using a steepest descent method to calculate moments of the bud size distribution (\eq{selfag}). We find $(\Delta R/R^*)^2=k_BT/(16\pi\kappa)B'_2/(B'_2-B_2^{eff})$, where $B'_2\equiv B_2+s_1/(2\phi^*(1-\phi^*))>0$ includes both the interaction between brushlets and the entropic contribution\footnote{if $B'_2<0$, the inclusions spontaneously demix on the flat membrane}. The mean radius variation shows the signature of the membrane curvature instability mentioned earlier\cite{leibler}. If the coupling between membrane and inclusion is sufficiently strong: $B_2^{eff}>B_2'$, or $D_{rr}^2>8k_BT\kappa B'_2$, then small fluctuations of any lengthscale are unstable and the mean variation of radius becomes large. The actual dispersion in bud size depends on how close we are to the instability. It is of order $6\%$ for the parameters used below. Note that variations in shape that conserve the mean curvature of the membrane should be larger, as they only cost a fraction of the energy penalty corresponding to variation of the bud global size.  From electron micrographs of caveolar membranes, the projected radius variation is of order $20\%$\cite{rothberg} (see also Fig.13.48 in \cite{alberts}).

Numerical calculation of the bud radius and protein concentration in the membrane is shown in Fig.3 upon variation of the surface tension for different values of the coupling strength (for an attractive energy $E_{att}=0.5\ k_BT$, see sec.\ref{results}). A strong variation in bud size is observed for small surface tension. At larger tension, the radius is almost insensitive to $\gamma$. Bud formation is however  less favorable, as can be seen from the increase of the c-b-c. Our model also predicts the existence of a critical point\cite{usbud}, hence a possible coexistence of buds of different radius, connected to the curvature instability studied by Leibler\cite{leibler}. We will not discuss this further here, as it is probably not relevant to the problem of caveolae formation.

We have derived the bud morphology as a function of two variables which depend on the actual shape of the force distribution: $B_2$ and $D_{rr}$. In order to make quantitative prediction, we
study below two ``extreme" force distributions.
 
\section{Results for various force distributions}\label{results}

In this section we discuss the results above for the polymer brush model and the block model.  The force distributions involve 3 parameters. The lengthscales $a$ and $b$ can be measured experimentally: $a\simeq 2nm$ and $b\simeq 5nm$. The energy scale of the force $E_0$ can be calculated for the brush model, and will be estimated for the block model(\eq{cc}). We present results for the minimal radius $R_{min}$ (connected to the force moment  $D_{rr}$) and the membrane mediated interaction $\Phi(r)$. The excluded volume  $B_2$  involves the full interaction potential and is discussed at the end of this section.

The calculation results in a unified description of the force distribution via the energy scale $E_0$ and the ratio of  size of the protein aggregate over deformation range $ka$ (for $ka\ll1$). The bud radius is $R_{min}/b=\alpha\kappa/E_0$ with
\be
\alpha_{block}={16 a\over b}\sim6\quad\alpha_{brush}={16\over 3}\left(2+{a\over b}\right)\sim 13
\ee
The oligomer-oligomer interaction is given by \eq{phisymm}, and involves the force moment $\zeta=-\beta(ka)^2 E_0/a$ with
\be
\beta_{block}={\pi b^2\over 8 a^2}=2.5\quad\beta_{brush}={3\pi b^2\over 8 a(a+2b)}=1.2\vspace{-0.5cm}
\ee

\vspace{0.5 cm}{\em Brush distribution}

For  ideal gaussian chains, the energy scale given by \eq{cc} is $E_0\simeq(Q/2\pi)^{3/2}\sim4k_BT$. Most of the force is concentrated near the center of the distribution, and has a small effect on the membrane. Fig.1 shows that the deformation is quite small (of order $0.4nm$). However, collective effects lead to the formation of fairly small buds of  minimum bud radius $R_{min}\simeq300 nm$ (much larger, however,  than the caveolae). The membrane mediated interaction between protein aggregates is very small, of order $10^{-3} k_BT$.  

\vspace{0.5 cm}{\em Block distribution}

The block distribution is probably more relevant to the case of stiff, short proteins such as caveolin. We choose the strength of the force so that each protein contribute to of order $k_BT$ of energy: $E_0\simeq10-15 k_BT$, which imposes a displacement $u(r=0)=2nm$ (Fig.1). The corresponding minimum bud radius $R_{min}\simeq 60nm$, is comparable to the radius of caveolae. Radius variation with surface tension is shown in Fig.3. The interaction potential is $\Phi(r)=0.02K_0(kr)\ k_BT$. We believe that, although small, this repulsive interaction might be responsible for the remarkable phase behavior of the proteins at the surface of the buds (sec.\ref{stripes}).

\vspace{0.5 cm}{\em Second virial coefficient and critical budding concentration}

The virial coefficient $B_2$ defined by
Eq.(\ref{pio6}) involves the full interaction potential $V(r)$,
including hard core repulsion and the
membrane mediated physical interaction \eq{phisymm}. Moreover, we have experimental evidence\cite{cterm1,structure2} that there exists short range specific attractions between
the protein side chains (C-termini). To describe this attraction, we
adopt the exponentially short range form
$V_{att}=-E_{att}e^{-(r-b)/b}$, where $E_{att}\sim k_BT$ is the strength of
the attraction, of range the size of the oligomer $b$. The short range attraction acts to increase the oligomer density inside the invagination. The resulting buds are crowded  with proteins $\phi\sim 0.8$, and are quite small, with a radius of order $R\sim 70nm$ - see Fig.3.

\section{Microphase separation at the bud surface}\label{stripes}

One peculiar
feature of the caveolae is their striated texture, believed to corresponds to alignment of protein oligomers at the
surface of these buds\cite{rothberg} (see also Fig.13.48 in \cite{alberts}). This finding is
particularly striking as it is not trivial to understand how radially
symmetrical oligomers may organize themselves into non-symmetrical phases. We argue
that the stripe phase might be a signature of the membrane mediated
repulsion between protein aggregates. Molecular dissection of the
caveolin protein has shown that the oligomers interact attractively via
the third distal region of their C-termini\cite{structure2}. This attraction may lead to gas/liquid phase separation of the caveolin oligomers, which results in dense membrane patches (the liquid) coexisting with
less dense
regions (the gas).  Our situation is more complex, as we have shown the
existence of an additional, membrane-mediated, longer range repulsion between  oligomers.
It has recently been argued at the light of computer  simulation\cite{stripes} that under this long range repulsion, the gas and liquid phases are broken into microdomains (circles at small concentration, and stripes for higher concentration). This is because large aggregates are costly
due of the long range repulsion, while small aggregates (circles
at low density, and stripes for larger density) are favored by the
short range attraction. A linear stability analysis for a solution of particle interacting with an exponentially short range attraction (of strength $E_a$ and range the particle radius $b$), added to the membrane mediated repulsion of \eq{phisymm}  (of strength $E_r$),  shows\cite{multipolepaper} that periodic
arrays of dense and dilute regions are expected for strong enough repulsion $E_r>\frac{3}{2}E_a(kb)^4$. Because of the long range of the repulsion (for oligomers, $b\simeq 5$nm, $k^{-1}\simeq 50$nm), a repulsive interaction as low as $10^{-2} k_BT$ between protein oligomers can indeed produce a well ordered phase for an attractive interaction of order $k_BT$. The structures have a typical size $2\pi/q^*$ which is, to lowest order, independent of the range of the repulsion:
$q^*b=(2E_r/3E_a)^{1/4}\simeq \frac{1}{4}$. It thus defines dense regions of size of order a few oligomer diameters, which compares quite well with the experimental observations.

\section{Conclusions}\label{concs}

We have constructed a theory for the formation of bud-like invaginations on fluid membranes, which we have compared with experimental data for caveolae found on the surfaces of cells. Our description incorporates the effect of both membrane rigidity and tension.

Proteins in the caveolin family are known to play a crucial role in the
formation of caveola, by forming homo-oligomers which concentrates in the buds. We argue that asymmetrically anchored membrane proteins (or protein oligomers for caveolin) can apply forces to the membrane. We examined
several models for the origin and magnitude of these forces, which may be purely entropic in origin or may result from stronger interactions. Such forces act to exert bending moments on the membrane and drive the formation of bud-like structures, for which we are able to make theoretical predictions. Our model correctly reproduces the size of the buds ($100$nm), and provides a physical explanation for the origin of the morphological striations observed on their surface. 

Our results also shed light on several experimental observations concerning the function of caveolae and the result of caveolin mutation. It has recently been suggested that caveolae-like domains play
a critical role in the mechanosensing and/or mechanosignal transduction of the
extracellular signal-regulated kinase pathway\cite{shear}. We predict that while the membrane tension has little effect on the size of the buds (Caveolae indeed have a similar structure in different kind of cell, which possibly bear different tension), the amount of caveolin protein required to observe bud formation does increase strongly with surface tension. Increasing the tension (either via a shear stress or by direct cell manipulation) may result in the disappearance of the buds if the available protein amount is insufficient. This is a testable prediction.

We have also described how the control of the morphology of caveolae can be
achieved in a number of ways, including by the level of cholesterol in a membrane. High cholesterol content in membrane is known to result in higher membrane rigidity, driving an increase in the bud radius.

Finally, caveolae formation in mutant caveolin
systems\cite{mutation} also provides verification of our theory. Mutants which lack the self-attractive segment of the N-termini (responsible for the formation of the homo-oligomers) are still competent to drive vesicle formation, but result in much larger buds $R\sim 1\mu m$. This is consistent with the fact that the force exerted by isolated proteins should be about 10 times smaller than the force exerted by oligomers resulting in a ten-fold increase of the bud radius. Mutants which lack the mutually attractive C-terminus result in similarly larger buds. Within our theory, this mutation results in a weaker oligomer-oligomer attraction, hence in a lower density of caveolin in caveolae and therefore larger buds. However, our theory predicts that oligomer attraction should not strongly influence the bud size. A natural conclusion would be that the C-termini contribute, either directly or indirectly, to the force exerted on the membrane, and that this force is reduced in mutants.

\noindent ACKNOWLEDGEMENTS. We gratefully acknowledge insightful
discussions with A. Johner, G. Rowland and R. Ball. One of us (MST) acknowledges the support of the Royal Society (UK) in the
form of a University Research Fellowship.

\newpage
\clearpage

\section*{Figure Captions}

FIG.~1. (a) Sketch of the blob model for the anchored protein aggregate and (b) force distribution for the two models used in the paper: brush distribution (dashed, corresponding to (a)) and block distribution (solid). The membrane is pushed down by the ``corona" of grafted polymers out to $r=b=5$nm and is pulled upwards by the anchored ``core" inside $r'=a=2$nm. (c) The corresponding membrane deformation $u(r)$ in unit $\frac{E_0}{\kappa}a$ for $k^{-1}=30nm$. The brush distribution has a weaker effect on the membrane because the force is mostly concentrated near its center ($r=0$). For aggregates residing on the cytoplasmic face of the membrane, including caveolin homo-oligomers the cell interior would be above the membrane.\\

\noindent FIG.~2.  Sketch of bud formation upon increase of oligomer concentration. (a)  Below the critical budding concentration (c-b-c), the membrane is uniformly covered by isolated oligomers. (b) At the c-b-c, buds have formed and outnumber isolated inclusions. (c) Above the c-b-c, the size and shape of the buds remains the same, and their number increases with the concentration.\\

\noindent FIG.~3.  Variation of the caveolae preferred radius $R^*$ (in nm) with the surface tension for 2 values of the coupling strength $E_0$ for a short range attraction of $E_{att}=0.5k_BT$ between brushlets (see sec.\ref{results}). Two inset curves show the variation of the bud composition $\phi^*$ (solid) and critical budding concentration (dashed) with for the same range of surface tension. The other parameters are $a=2nm$, $b=5nm$, $\kappa=20k_BT$

\newpage

\clearpage

\pagestyle{empty}

\begin{figure}
\vskip 3truecm
\centerline{\hspace{1 cm}\epsfig{file=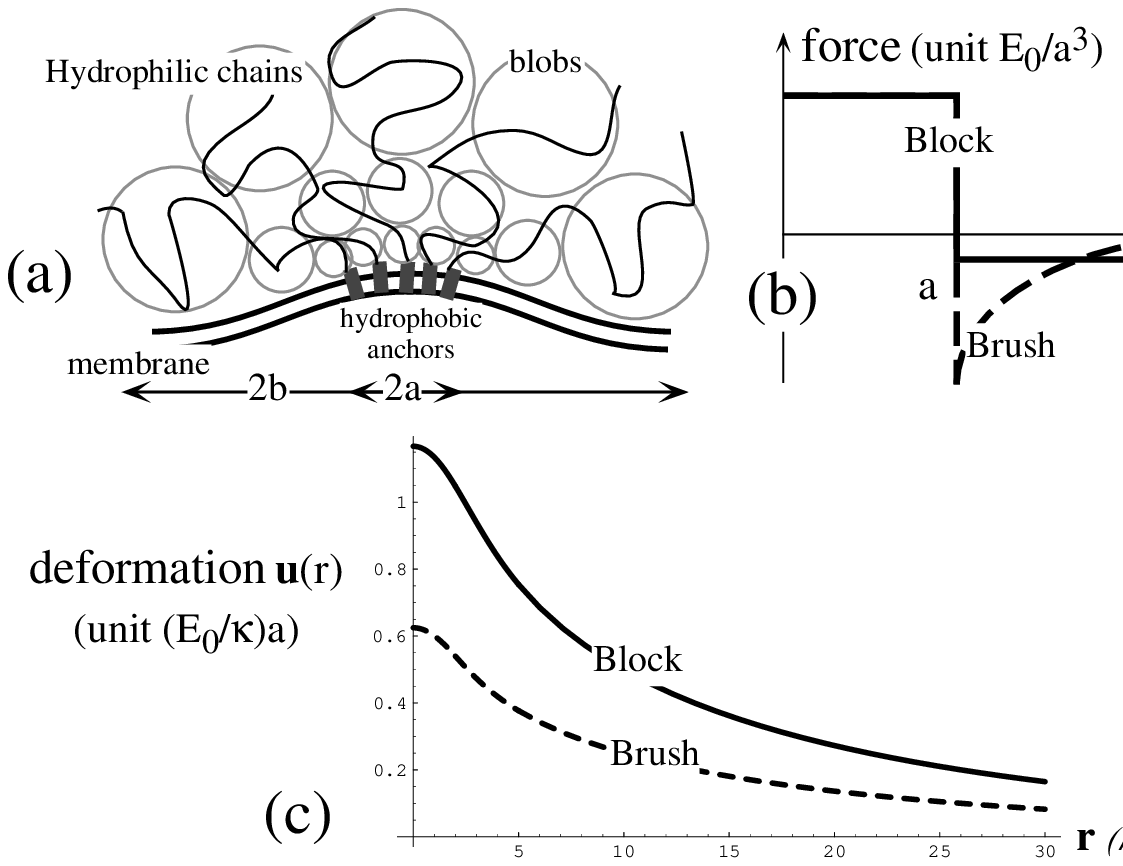,height=7cm}}
\end{figure}

\begin{center}
~\\~\\~\\~\\~\\~\\~\\~\\Figure 1 - Sens \& Turner
\end{center}

\newpage
\clearpage

\pagestyle{empty}
\begin{figure}
\vskip 3truecm
\centerline{\hspace{1 cm}\epsfig{file=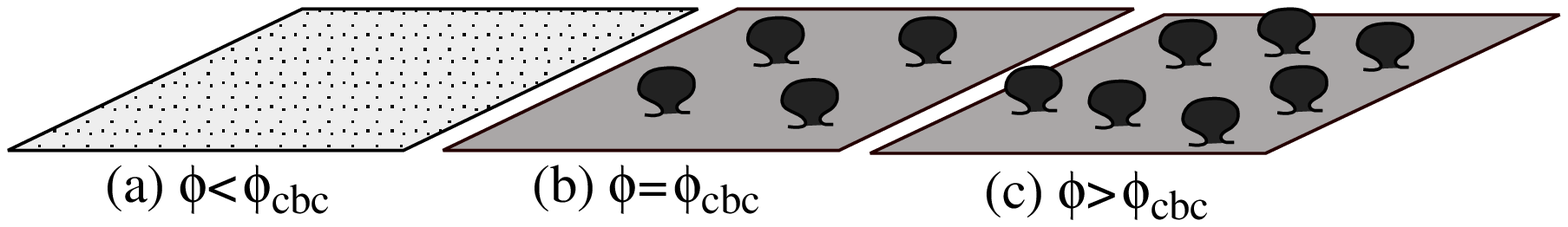,height=2cm}}
\end{figure}

\begin{center}
~\\~\\~\\~\\~\\~\\~\\~\\Figure 2 - Sens \& Turner
\end{center}

\newpage
\clearpage
\pagestyle{empty}
\begin{figure}
\vskip 3truecm
\centerline{\hspace{1 cm}\epsfig{file=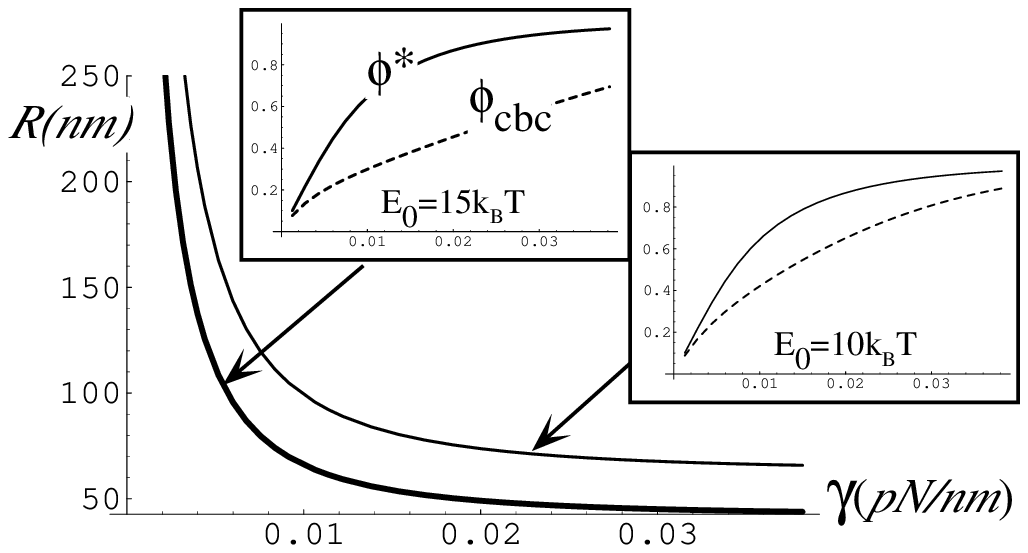,height=6cm}}
\end{figure}

\begin{center}
~\\~\\~\\~\\~\\~\\~\\~\\Figure 3 -Sens \& Turner
\end{center}


\begin{thebibliography}{xx}
\bibitem{nonetforce} No net force acts on the membrane, as no net force can be exerted by an object that is not externally attached (i.e to the cytoskeleton). For more details
see Ref.\cite{multipolepaper}
\end{thebibliography}
\end{document}